\documentclass[review]{elsarticle}

\usepackage{hyperref}
\usepackage{amsmath,amssymb,amstext,amsthm,tcolorbox}
\usepackage{graphicx}
\usepackage[utf8]{inputenc}
\usepackage[left=2cm,right=2cm,top=1cm,bottom=3cm]{geometry} 
\usepackage{amsmath}
\usepackage{array}
\usepackage{hyperref}
\usepackage{float}
\usepackage{soul}
\usepackage{multirow}
\theoremstyle{plain}
\hypersetup{
	colorlinks=true,
	linkcolor=blue,
	filecolor=magenta,      
	urlcolor=cyan,
}

\usepackage[ruled,vlined]{algorithm2e}   

\newcolumntype{P}[1]{>{\centering\arraybackslash}p{#1}}

\usepackage{setspace,color}   

\newtheorem*{theorem*}{Theorem}

\usepackage{bbm,lipsum}

\numberwithin{theorem}{section}

\journal{Journal of \LaTeX\ Templates}

\begin{document}
	
	\begin{frontmatter}
		
		\title{Considering user dynamic preferences for mitigating negative effects of long tail in recommender systems }
		
		
		\author[mymainaddress]{Reza Shafiloo}
		\ead{rezashafilou@eng.ui.ac.ir}
		\author[mymainaddress]{Marjan Kaedi\corref{mycorrespondingauthor}}
		\ead{kaedi@eng.ui.ac.ir}
		\author[mysecondaryaddress]{Ali Pourmiri}
		\ead{ali.pourmiri@mq.edu.au}
		\cortext[mycorrespondingauthor]{Corresponding authors.}
		
		\address[mymainaddress]{Faculty of Computer Engineering, University of Isfahan, Azadi Sq., Hezarjarib St., Isfahan, Iran}
		\address[mysecondaryaddress]{Macquarie University, Sydney, Australia}
		
	\begin{abstract}
		\noindent Nowadays, with the increase in the amount of information generated in the webspace, many web service providers try to use recommender systems to personalize their services and make accessing the content convenient. Recommender systems that only try to increase the accuracy (i.e., the similarity of items to users' interest) will face the long tail problem. It means that popular items called short heads appear in the recommendation lists more than others since they have many ratings. However, unpopular items called long-tail items are used less than popular ones as they reduce accuracy. Other studies that solve the long-tail problem consider users' interests constant while their preferences change over time. We suggest that users' dynamic preferences should be taken into account to prevent the loss of accuracy when we use long-tail items in recommendation lists. This study shows that the two reasons lie in the following: 1) Users rate for different proportions of popular and unpopular items over time. 2) Users of all ages have various interests in popular and unpopular items. As a result, recommendation lists can be created over time with a different portion of long-tail and short-head items. Besides, we predict the age of users based on item ratings to use more long-tail items. The results show that by considering these two reasons, the accuracy of recommendation lists reaches 91\%. At the same time, the long tail problem is better improved than other related research and provides better diversity in recommendation lists in the long run. 
	\end{abstract}
	
	
\end{frontmatter}

\section{Introduction}
These days, with the development of information technology, human beings are more inclined to use Internet-based systems such as education systems, e-commerce, and social networks. Hence the volume of content produced in these systems is increasing dramatically. Therefore, the services provided in these systems are personalized by recommender systems based on users' interests to give them a better experience of working with these systems. Also, they separate unrelated items from the mass of information by recommending content and items appropriate to users' interests \cite{wang2016multi,zhou2021generic,deldjoo2021explaining}. 
Recommendation systems that seek to increase only the accuracy in the recommendation lists are called accuracy-based recommendation systems, which cause the long tail problem in recommender systems \cite{hamedani2019recommending,gharahighehi2021fair,boratto2021connecting}. This problem arises when accuracy-based recommendation systems suggest more popular items with more rates \cite{zuo2015personalized,hamedani2019recommending}. Therefore, many items rated by fewer users are not included in the recommendation lists and left out in the long tail. However, a small part of the items called short head items that are rated by more users constantly appear in the recommendation lists, and this problem will be exacerbated in the future \cite{park2008long,wang2016multi}.
The main problem regarding the lack of long-tail items in recommendation lists is that sales of many items decrease and become more acute in the long term. As the long tail problem in recommender systems gradually duplicates popular items for users, the working experience with the system becomes boring \cite{jain2020multi,berbague2021overlapping}. Several studies have been conducted on the diversification of recommendation lists using lesser-known items that belong to long tail to solve the long tail problem in recommender systems \cite{jain2020multi,wang2016multi,hamedani2019recommending}. Therefore, for better performance of recommender systems, both the goal of "solving the long tail problem" and "accuracy" must be achieved. Nevertheless, these two goals are usually in trade-off since using the long tail items reduces the recommendations' accuracy \cite{wang2016multi,hamedani2019recommending,cui2017novel}.

On the other hand, in the design of most recommender systems, it is often assumed that the pattern of user interactions with the system does not change over time and users' preferences and interests are constant over time \cite{zheng2018tag,wang2021attention}. However, research on user preferences has shown that users' interests and does not remain constant over time \cite{li2019sparse,liu2018learning,pereira2018analyzing,bagher2017user}. Considering users' dynamics preferences increases the accuracy of recommendation lists and prevents items that are not related to the users' current interests from recommending \cite{zheng2018tourism,wang2021attention}. Therefore, it is essential to consider changes in user preferences over time to create recommendation lists that are more in compliance with current user preferences \cite{zheng2018tag}.

To our knowledge, in all the studies done on the long tail problem and the diversification of recommendation lists based on user preferences, the preferences and interests of users over time have been considered consistently and unchanged. However, as mentioned, user preferences are not fixed and change over time. In this study, the intention is to solve the long tail problem in recommender systems by considering the dynamic preferences of users. By considering the dynamic preferences of users, the accuracy of the recommendation lists increases, and therefore more items from the long tail can be used with less accuracy loss in the recommendation list.
\section{Literature review}
As mentioned earlier in this paper, we want to solve the long tail problem by considering users' dynamic preferences. Therefore, our literature review has two main subsections. First, we review studies related to solving the long tail problem and then studies related to the user's dynamic preferences.
\subsection{Long tail problem}
In 2008, park et al. found that items that received fewer ratings or were recently added to the recommender systems participate less in recommendation lists. They predict long-tail items' ratings for users by clustering techniques \cite{park2008long}.

Malekzadeh and Kaedi present a method that maintains accuracy by personalizing the diversity of recommendation items \cite{hamedani2019recommending}. In this method, a multi-objective optimization problem is defined that meets the goals of long-tail item participation, diversification, and accuracy. 

In another study, Alshammari et al. propose a different approach that uses collaborative and content-based algorithms together to solve the long-tail problem and improve the prediction error of long-tail item ratings \cite{alshammari2017hybrid}. 

Using users' sentiments about the items based on posted comments is another approach to solving the long tail problem \cite{huang2019novel}. Huang et al. try to increase the accuracy of finding similar items by extracting information from the recorded comments for the items.

In another study, the Siamese network (a machine learning method for a limited data set) solved a long tail problem \cite{sreepada2020mitigating}. Building rating predictor models based on neural networks requires much data to have high prediction accuracy. However, the number of ratings recorded for long-tail items is minimal. Therefore, Sreepada et al. use Siamese networks to create a suitable model with insufficient training data. 

In another research approach, multi-objective evolutionary optimization algorithms try to solve the long tail problem \cite{wang2016multi}. Wang et al. define two objective functions to solve the long tail problem: The first one calculates the accuracy of lists, and the second objective function reduces participation of popular items.

In another approach, instead of using the usual methods of recommender systems, methods based on multi-component graphs are used to create recommendation lists \cite{luke2018recommending}. Luke et al. place items, genres, and users in different sections and draw edges based on rating history. Then they traverse the graph using the Markov process. In this method, the probability of selecting items with a lower number of ratings and a bigger average rating is high for traversing.

Karakaya and Aytekin proposed a method based on transposing items graph increase diversity in the recommendations list \cite{karakaya2018effective}. Each node represents an item in the graph model, and the edge between the two nodes indicates that the two items co-occur in the recommendation lists. The weight of the edges also indicates the number of repetitions of the two items in the recommendation lists. Then with a prospect, the recommendation list items are replaced with the graphic model items. To solve the long tail problem, the items with a higher number of ratings are less likely to be replaced with other items.
These studies solve the long-tail problem with different approaches; however, they consider users' interests constant. As we review in the next section, user preferences change over time, and we can consider users' interests in long-tail items dynamic.
\subsection{User's dynamic preferences} 
Some studies have different methods for considering users' dynamic preferences in recommender systems. Liu et al. set a time frame for all users that reflects their recent interests.  In this study, items are likely to be used in the recommendation list based on categories of items that users have recently rated \cite{liu2018learning}.

Pereira et al. identify users' dynamic preferences based on their preferences graph and their change in position in the social network graph \cite{pereira2018analyzing}. In this method, first, all user preferences are extracted over time. If a variance is found between the user preferences, the preferences change will be detected, and the user preferences graph will be updated. After updating the preferences graph, the user's social network graph is checked, and changes in the user's social network position and preferences graph are examined.

Rahimpour et al. identify dynamic user preferences by modeling users based on the trends over time. In this method, They cluster users first. Then, the user rate prediction model is then built based on trends of items in the cluster. A user-generated model shows how likely a user is to be interested in popular items in that cluster. In this study, clusters can be changed over time. Therefore, dynamic users' preferences are modeled using trends, and dynamic clustering \cite{bagher2017user}.

In another study to model the dynamic preferences of users, a network of users is constructed according to their relationships to create recommendation lists for users, and this graph changes over time \cite{rezaeimehr2018tcars}. Rezaeimehr et al. extract time-based association rules from the users' community and create recommendation lists accordingly.

Zheng et al. use users' feelings and changes in their preferences to predict better users' next travel destination in the travel recommender system \cite{zheng2018tourism}. In this method, two time intervals are considered as the most significant part of preferences changes. The first interval is related to the seasons, and the next is related to the holidays. Zheng et al. consider users' travel preferences according to these intervals.

In another study, Li et al. consider factors for the collaborative filtering algorithm to use the dynamics of users' preferences for creating recommendation lists \cite{li2019sparse}. Li et al. predict the rating of items by considering the trends and users' preferences. In this method, based on the user's activities, some users are considered similar users. If a user registers a new rating for an item, the list of similar users will be updated based on his recent activities, and Li et al. predict the rating based on new similar users.
Although users' dynamic preferences are used to recommend accurately, they are not used to solve the long tail problem. Also, we will consider users' age changes for solving this problem better.
\section{Proposed method}
\begin{figure}[t]
\begin{minipage}{1\textwidth}
	\centering
	\includegraphics[scale=0.53]{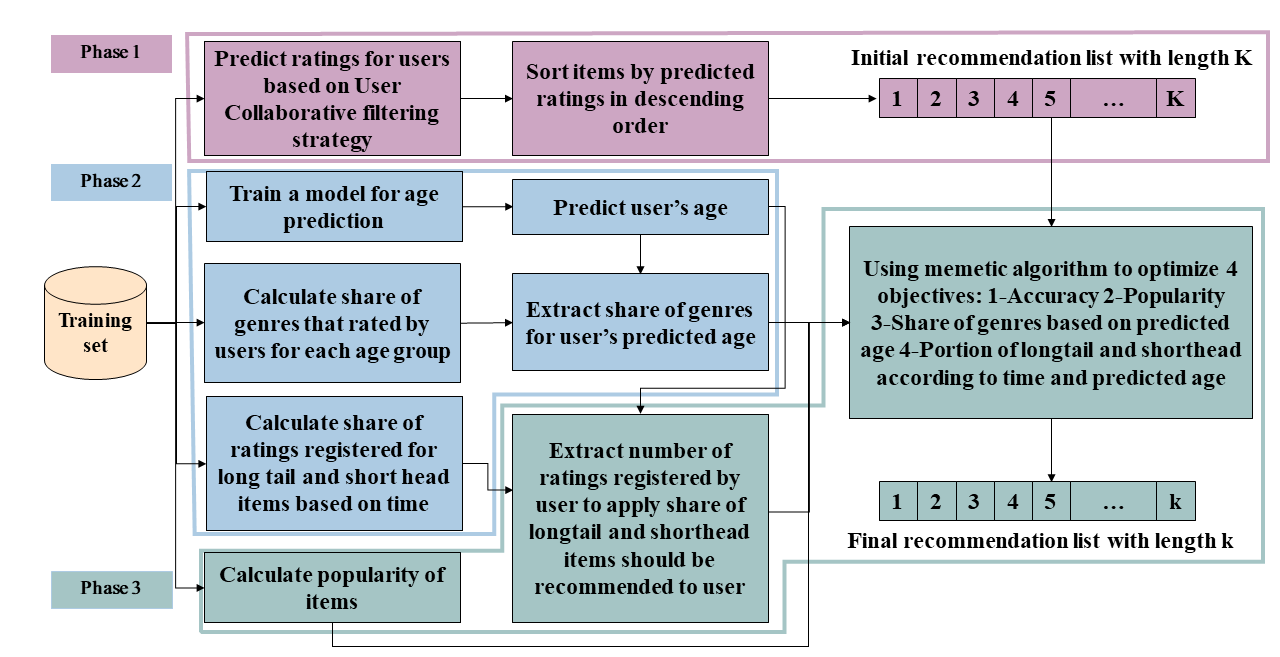}
	\caption{ Summary of the proposed method steps. }\label{Fig:flowchart}
\end{minipage}\hfill
\end{figure}

This section introduces the proposed method that solves the long tail problem by considering the users' dynamic preferences. Users show different interests in different genres according to their age. Also, The level of interest at different ages in the long tail and short head items change over time. Therefore, these factors are used to achieve better results. We generally divide the proposed method into three steps. In the first step, one of the fundamental algorithms of the recommender systems predicts the ratings that users have not registered for the items. In the next step, age prediction is performed using age prediction models obtained using machine learning methods.  Finally, we optimize the recommendation lists according to the objectives of the problem by using one of the multi-objective evolutionary optimization algorithms called the Memetic algorithm. All the innovations of this study are related to steps 2 and 3. Step 1 used the fundamental algorithms of recommender systems. To solve this problem, we choose the movieLens dataset to evaluate our work. Nevertheless, we can use the proposed method on other datasets that have the long-tail problem. The figure \ref{Fig:flowchart} summarizes all the steps.
\subsection{Step 1: Predicting item ratings}
The first step predicts the ratings of unknown items for users. Various algorithms related to recommender systems can be helpful for this purpose. We use the user-based collaborative filtering algorithm to find similar users and predict item ratings \cite{adomavicius2011improving,yue2021optimally} since, in another part of this work, we will find users of the same age and recommend new items based on users of the same age interests.
The similarity of the two users $u$ and $v$ is shown in equation \ref{sim}.
\begin{align}\label{sim}
sim(u,v)=\frac{\sum_{k \in I_u\cap I_v} (r_{uk}-\mu_u).(r_{vk}-\mu_v)}{\sqrt{\sum_{k \in I_u\cap I_v} (r_{uk}-\mu_u)}.\sqrt{\sum_{k \in I_u\cap I_v} (r_{vk}-\mu_v)}}
\end{align}
$I_u$ is a subset of items rated by user $u$. Also, $r_{uk}$ indicates the user's ratings on item $k$, and $\mu_u$ indicates the average rating registered by user $u$. In the following equation, $\hat{r}_{uj}$ shows the predicted rating of user $u$ to the item $j$, which is calculated as follows.
\begin{align}\label{Pr}
\hat{r}_{uj}=\mu_u+\frac{\sum_{v \in p_u(j)} sim(u,v).(r_{vj}-\mu_v)}{\sum_{v \in p_u(j)} |sim(u,v)|}
\end{align}
In equation \ref{Pr}, $P_u(j)$ is a set of closest users to the target user $u$ that have registered ratings for item $j$.
\subsection{Step 2: Predicting the user's age to be used in building recommendation lists}
Age prediction plays an important in personalizing and provides valuable information for users' preferences \cite{pandya2020use}. In the second step, we predict the user's age by using machine learning algorithms according to the user's preferences for items. Therefore, first, we explain what methods and data we use to model the user's age. Then, we describe the application of age prediction in this research.
\subsubsection{user's age prediction}
As mentioned before, we evaluate our work by the movielens dataset. For age prediction, we use the method proposed in our previous study that used user rating. \cite{shafiloo2021predicting}. We used the MovieLens dataset for age prediction and built a user model based on genres that users rated for. Also, we used two types of information presented for movies in the IMDB database: Parental Guide Information and Motion Picture Association of America. Results have shown that using user ratings for movies can acquire an acceptable level of accuracy for age prediction compared with other methods. 
\subsubsection{The relationship between age and preferences}
\begin{figure}[t]
\begin{minipage}{1\textwidth}
	\centering
	\includegraphics[scale=0.53]{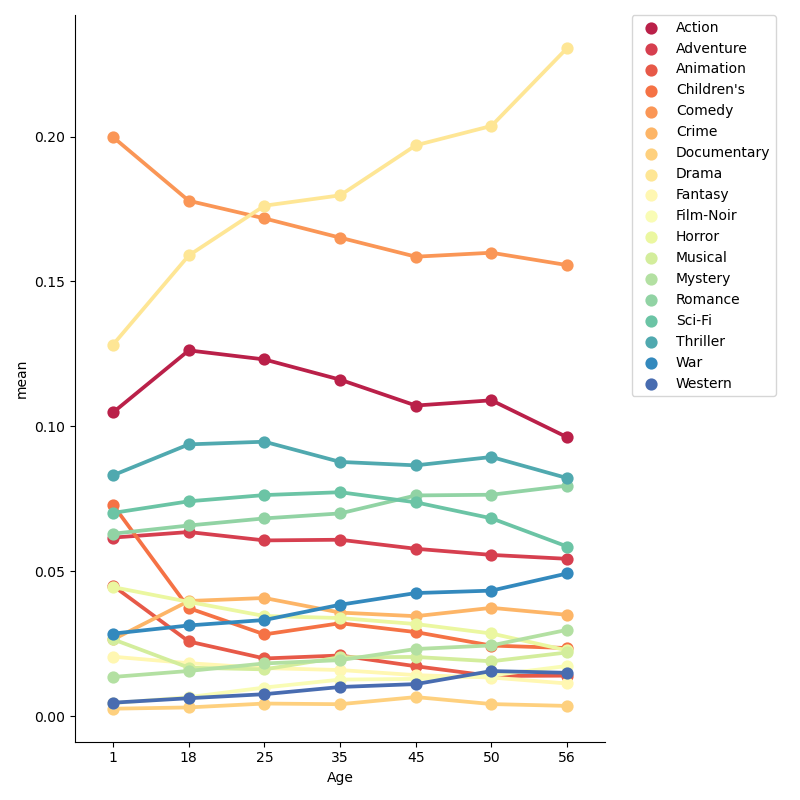}
	\caption{Mean of the portion of users' rating relating to different genres for all items in 7 different age group}\label{Fig:UserGenreAllData}
\end{minipage}\hfill
\end{figure}

One of the contributions in this study is that we use users' age and their interests to build recommender systems. In this study, we consider users of different ages and their preferences for various genres. We calculate the share of genres rated by users based on their age. Figure \ref{Fig:UserGenreAllData} shows the average ratings registered by users of different ages for various genres in the MovieLens dataset. This figure illustrates that users have particular interests at different ages. Thus, we can use long-tail items in the recommendation lists based on users of the same age's interest in different genres.
\subsubsection{Changes in user preferences in the long tail and short head items over time}
\begin{figure}[t]
\begin{minipage}{1\textwidth}
	\centering
	\includegraphics[scale=0.7]{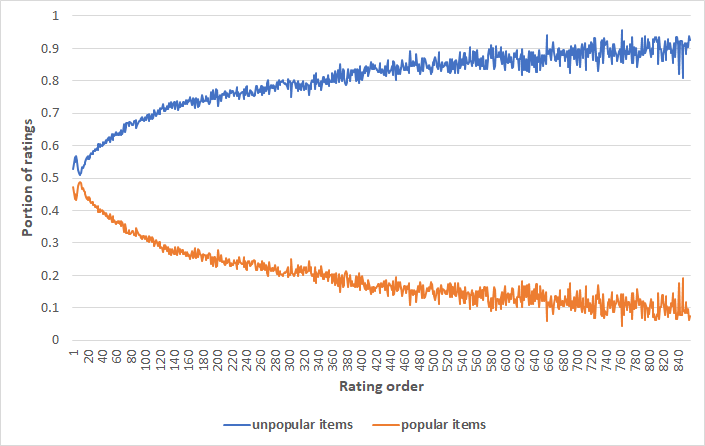}
	\caption{Mean of the portion of users' rating relating to different genres for all items in 7 different age group}\label{Fig:portionShorthead}
\end{minipage}\hfill
\end{figure}

In this study, by investigating the ratings that users have registered for long tail and short head items, it is concluded that users' preferences in the long tail and short head items change over time , and figure \ref{Fig:portionShorthead} shows this fact.

\begin{figure}[t]
\begin{minipage}{1\textwidth}
	\centering
	\includegraphics[scale=0.4]{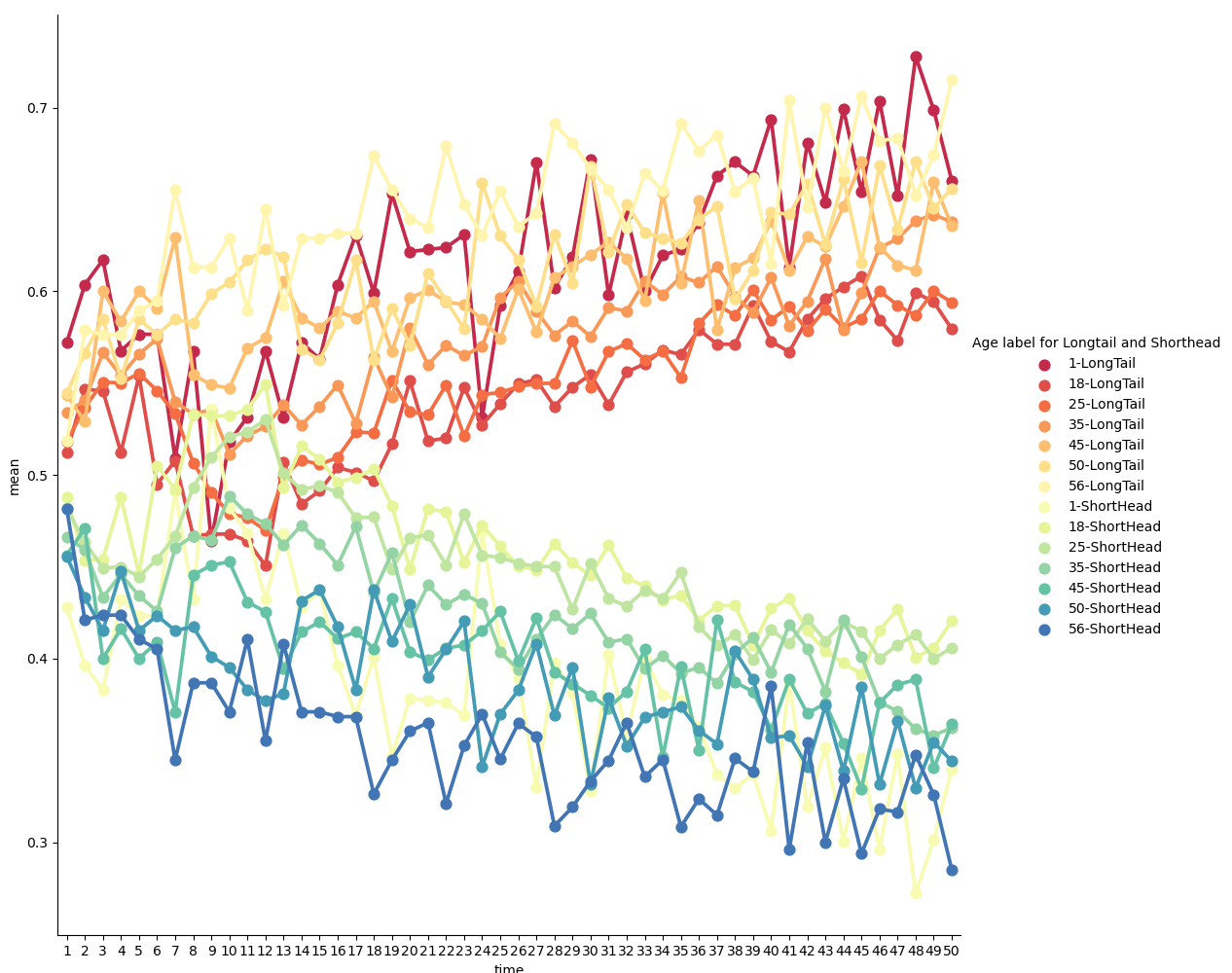}
	\caption{share of ratings that users with different ages over time scored for long tail and short head items}\label{Fig:ageLSTime}
\end{minipage}\hfill
\end{figure}

As shown in Figure \ref{Fig:portionShorthead}, users' interest in the long tail and short head items is almost equal in their early ratings. However, over time, users prefer to visit long-tail items and rate them. Next, we investigate the level of users' interest in the long tail and short head items by age over time. Figure \ref{Fig:ageLSTime} shows the share of ratings that users with different ages rated for long tail and short head items over time.
According to figure \ref{Fig:ageLSTime}, it is clear that users' level of interest at different ages in the long tail and short head items varies. For example, in figure \ref{Fig:ageLSTime}, 56-year-old users lose interest in short-head items much faster than other users and start rating for long-tail items in a shorter period. On the other hand, 18-year-olds lose interest in short head items less quickly than older people over time. 

According to the above explanations, the recommendation lists can be constructed by using the suitable ratio of the long tail and short head items according to the predicted age of the user.
\subsection{Step 3: Recommendation lists optimization}
In this study, four objectives for optimization have been considered to solve the long tail problem by considering the users' dynamic preferences. In the rest of this section, we describe the objectives intend to optimize lists.
\subsubsection{First objective: Reducing the number of popular items in recommendation lists}
In this study, the most crucial goal is to solve the long tail problem in the recommendation lists. Therefore, less popular items should be used in the recommendation lists to increase the participation of long-tail items in the recommendation lists. In this study, the popularity of items is defined according to the number of ratings registered for them. Therefore, the first objective is the increasing participation of long-tail items, defined as Equation \ref{LTpr} \cite{hamedani2019recommending}.
\begin{align}\label{LTpr}
Long\,tail\,item\,participation=\sum_{i=1}^{k} popularity(i)
\end{align}
In this equation, k indicates the length of the recommendation list. The smaller the values obtained from this equation, the less popular items and more long-tail items are used in the recommendation list.
\subsubsection{Second objective: Increasing the accuracy of recommendation lists}
In studies that attempt to increase the diversity of recommendation lists, they often experience a decline in accuracy because diversity and accuracy are in the trade-off. Therefore, to not reduce the accuracy of the lists while improving the diversity of recommendation lists, one of the optimization objectives is to increase the accuracy of the recommendation lists. This objective is defined as equation \ref{Acc} \cite{hamedani2019recommending}.
\begin{align}\label{Acc}
Accuracy=\frac{1}{\sum_{i=1}^{k} \hat{r}_{i}}
\end{align}
In the Equation \ref{Acc}, the smaller value obtained shows that $\hat{r}_{i}$ predicted for the items on the recommendation list is higher.
\subsubsection{Third objective: Considering user's dynamic preferences}
As explained, users' interest in the long tail and short head items changes over time, and users become more interested in registering ratings for long-tail items in the long term. Therefore, over time, recommendation lists can be created with more long-tail items. This study calculates the amount of user activity and the number of ratings recorded for the items over time. Then, according to the number of ratings that the user registered in recommender systems and user age, we consider his interest in long-tail items in the current time to his recommendation lists, which we define it as Equation \ref{LTRec}.
\begin{align}\label{LTRec}
\begin{matrix}
	Number\,of\,long\,tail\,items\,in\,list\,based\,on\,user\,age=|number\,of\,long\,tail\, items\,in\,current\,list \\
	- number\,of\,long\,tail\,items\,based\,on\,number\,of\,ratings\,registered\,and\,user\, predicted\,age|
\end{matrix}
\end{align}
\subsubsection{Fourth objective: Participation of long-tail items by considering their genre}
Users usually register fewer ratings for long-tail items. For this reason, it is necessary to use other information to find long-tail items closer to the user's preferences. As mentioned, in this study, we use the age of users to increase the accuracy of recommendation lists. Therefore, we can use the user's predicted age and favorable items to other users of the same age. As mentioned in the previous sections, users have a particular interest in different genres. Therefore, we can personalize the genres in the target user's recommendation list according to his age. We formulate the fourth objective as follows.
\begin{align}\label{GAge}
\begin{matrix}
	Genres\,based\,on\,user\,age=\\
	|PGU1-PGL1|+...+|PGU18-PGL18|
\end{matrix}
\end{align}
Equation \ref{GAge} shows the distance between the ratio of each genre in the recommendation list (PGL) and the ratio of that genre rated by users of the same age (PGU). The lower values obtained from this equation mean that the list created for the target user is close to the interest of other users at the same age. In this way, long-tail items can be included in the recommendation lists according to their genres.
\subsection{Optimizing recommendation lists using Memetic algorithm}
As the other studies\cite{karabadji2018improving,hamedani2019recommending,wang2016multi}, we could use multi-objective optimization algorithms to optimize recommendation lists. Other studies used different algorithms to solve the long tail problem. In this study, to best of our knowledge, it is the first time memetic algorithm is used for solving the long-tail problem.
\begin{figure}[t]
\begin{minipage}{1\textwidth}
	\centering
	\includegraphics[scale=0.6]{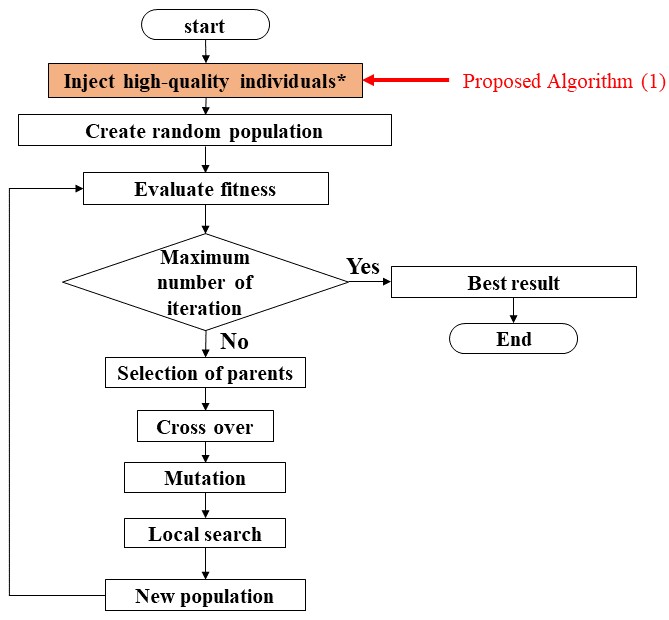}
	\caption{Problem optimization diagram using memetic algorithm}\label{Fig:memetic}
\end{minipage}\hfill
\end{figure}
As shown in Figure \ref{Fig:memetic}, many parts of the memetic algorithm are similar to the basic genetic algorithms. However, in the memetic algorithm, after crossover and mutation, a local search is performed. If a better solution is found in the local search, this solution is replaced in the created population \cite{sun2019interval}. In this study, in order to obtain better recommendation lists, the initial population is improved by using items that produce better results according to the objectives of the problem. We explain the proposed method in the next section.
\subsubsection{Proposed algorithm to improve the performance of the memetic algorithm}
In this study, to obtain better recommendation lists that improve results, the initial population is improved by using items cause better results according to the objectives of the problem. Algorithm \ref{Alg1} Add higher quality items.
\begin{algorithm}\label{Alg1}
\KwIn{Bag: a bag of Items}
\KwOut{Output bag with P number of Items}
\While(repeat this to find P unique Items){$i < P$}{
	$Chosen Item \longleftarrow$ choose a random item from Bag\;
	$Random Number \longleftarrow$ create a random number between [0,1]\;
	$Possibility \longleftarrow$ calculate possibility of Chosen item according to Equation \ref{eqPosibility}\;
	\If{$Random Number < Possibility$}{
		Add Chosen Item to output bag\;
		Remove Chosen Item from Bag\;
	}\caption{Injecting Items to improve diversity and accuracy}
}
\end{algorithm}

As shown in Algorithm \ref{Alg1}, a bag of items is given to this algorithm as an input, and then some items are randomly selected. A random number is then generated for each item selected. Next, by using equation \ref{eqPosibility}, the probability of choosing this item is calculated. If the calculated possibility is higher than the random number, we can add this item to the output bag.
\begin{align}\label{eqPosibility}
Possibility=\frac{M_i.e^{-a*N_i}}{5}
\end{align}
In Equation \ref{eqPosibility}, $i$ shows indices of chosen item, and $M_i$ represents the average ratings of users of the same age for the item $i$. $N_i$ also indicates the number of repetitions of the item $i$ in the previous recommendation lists. Therefore, the higher the value of $N_i$ depicts that this item has been recommended in the last recommendation lists. Consequently, the value of $e^{-a*N_i}$ decreases, and the probability of its presence in the following recommendation lists will reduce, and items that were not in the previous recommendation lists will be more likely to be included in the following recommendation lists. Finally, since the value of $e^{-a*N_i}$ is a number between zero and one, and the scores have values between [0,5], to normalize, the total value obtained is divided by 5.

Due to the increasing value of $N_i$ over time, the output value of function $e^{-a*N_i}$ decreases. Therefore, the probability of attendance short head items with lots of ratings decreases over time. This opportunity will be given to long-tail items with a lower number of ratings.
\section{Results and discussion}
This section first introduces the datasets used to evaluate the proposed method. Then, we introduce the algorithms used in this study for the comparison of the results. Next, the genetic algorithm's parameters and the proposed algorithm's parameters are introduced. Finally, we represent the evaluation criteria and the comparison results.
\subsection{Datasets}

This paper uses the Movielens dataset, commonly used to evaluate proposed methods of solving long-tail problems in other studies. The Movielens dataset has various versions \cite{noauthor_movielens_2013}. We use MovieLens 1M dataset that has 6040 users and 1 million ratings for 3883 items. Also, for acquiring more information about movies, we use IMDB Dataset. From this dataset, we use two types of information: Parental Guide \cite{imdbp}, and Motion picture association system \cite{motion}. The motion picture association system shows that how much content of a movie is suitable for children. Parental Guide measures the content of movies by 5 factors: (sex \& nudity), (violence \& gore, profanity), (alcohol \& drug \& smoking), and (frightening \& intense scenes).
\subsection{Algorithms used for comparison}
Several algorithms have been used for comparison to demonstrate the better performance of the method presented in this study. In general, we can divide the algorithms used for comparison into two general categories. The first category is the baseline algorithms of recommendation systems. In the second category, baseline algorithms are first used; then, optimization methods are used to provide better results according to the problem. We choose algorithms in the second category that solve the long tail items problem. In the following, we will explain these two categories of algorithms.
\subsubsection{First category: baseline algorithms of recommendation systems}
Item-based collaborative filtering algorithm is one of the most widely used recommendation algorithms. In this algorithm, the similarity of items is first calculated. The item rating is then calculated based on how similar the items are to each other. Finally, the items with the highest predicted ratings are suggested to the target user \cite{wang2017hybrid,chen2020collaborative}. Another baseline algorithm used to compare results is the user-based collaborative filtering algorithm. This algorithm first finds similar users and then predicts item ratings based on similar users' registered ratings for the same item \cite{sanchez2019building,iwanaga2019improving}.
\subsubsection{Second category: multi-objective optimization algorithms used to solve the long tail problem}
Among all the reviewed studies, two studies that have used multi-objective evolutionary optimization methods to solve the long tail problem have been compared with this study. Malekzadeh and Kaedi \cite{hamedani2019recommending} use the simulated annealing algorithm to solve the long-tail problem. Also, Wang et al. \cite{wang2016multi} use the genetic algorithm to solve this problem. As explained in section 3, the memetic algorithm is used for multi-objective optimization to solve the long sequence problem in this study.
\subsection{Evaluation metrics}
Studies mitigating the long tail problem have considered different criteria for evaluation. In this study, three criteria have been considered to compare the results.For accuracy metric, we choose precision critrion is defined as \ref{Precision} \cite{wang2016multi}.
\begin{align}\label{Precision}
Precision=\frac{N_{rs}}{N_{s}}
\end{align}
$N_{s}$ is the total number of items recommended to the user. $N_{rs}$ are also relevant items suggested to the user. Relevant items receive ratings that are higher than the user's average ratings.

The following criterion is aggregate diversity, which takes into account the number of items offered to users. This measurement counts how many the proposed algorithm can recommend different types of movies, especially long-tail items in the recommendation lists \cite{hamedani2019recommending}.
The last criterion is Novelty, which is calculated as \ref{Novelty} \cite{hamedani2019recommending}.
\begin{align}\label{Novelty}
Novelty=\frac{1}{\sum_{all\,recommended\,items} Popularity(items)}
\end{align}
This equation indicates that the more popular the items, the greater the denominator of the fraction per recommendation list, and therefore the lower novelty of the recommendation list.
\subsection{Results and Discussion}
\begin{figure}[h]
\begin{minipage}{0.4\textwidth}
	\centering
	\includegraphics[scale=0.38]{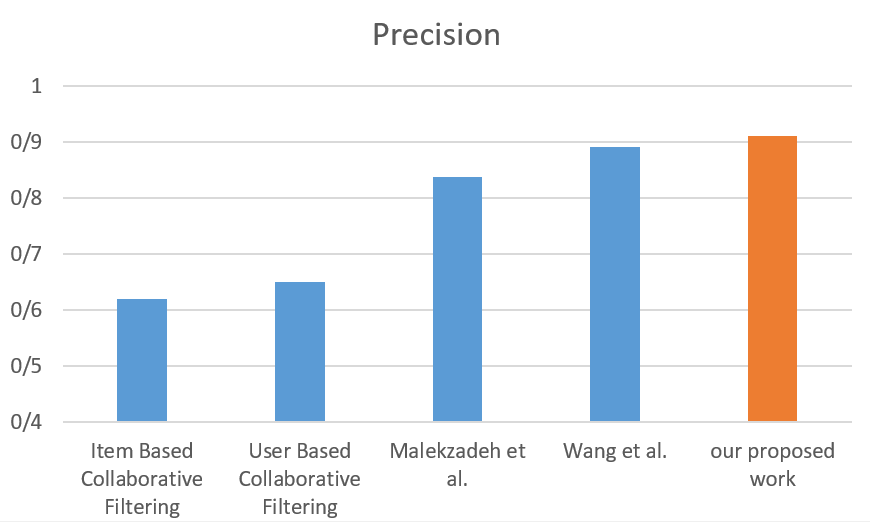}
	\caption{Comparison of precision criteria for Malekzadeh and Kaedi \cite{hamedani2019recommending}, Wang et al. \cite{wang2016multi} researches and user-based and item-based collaborative filtering algorithms and our proposed method}\label{Fig:precision}
\end{minipage}\hfill
\begin{minipage}{0.4\textwidth}
	\centering
	\includegraphics[scale=0.38]{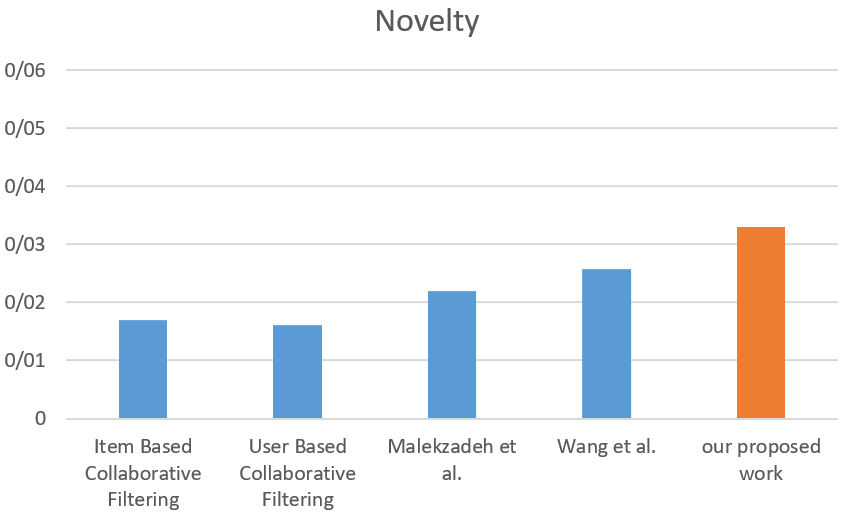}
	\caption{Comparison of novelty criteria for Malekzadeh and Kaedi \cite{hamedani2019recommending}, Wang et al. \cite{wang2016multi} researches and user-based and item-based collaborative filtering algorithms and our proposed method}\label{Fig:novelty}
\end{minipage}\hfill
\begin{center}
	\begin{minipage}{0.5\textwidth}
		\centering
		
		\includegraphics[scale=0.4]{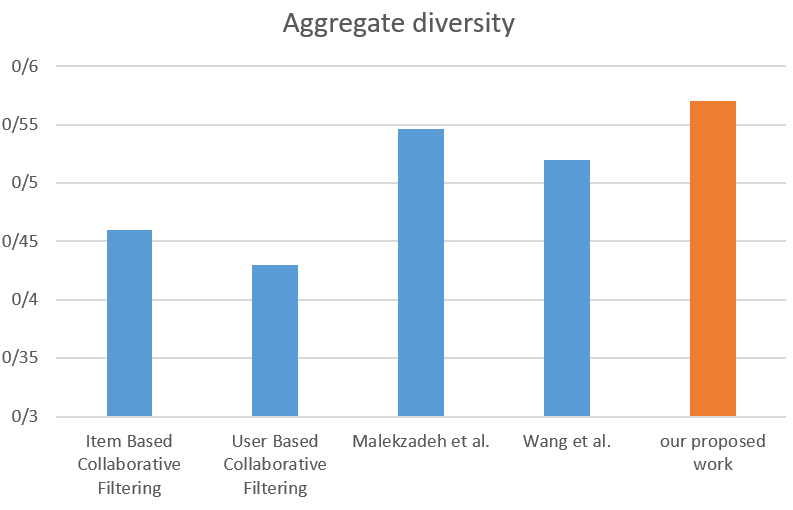}
		\caption{Comparison of Aggregate Diversity criteria for Malekzadeh and Kaedi \cite{hamedani2019recommending}, Wang et al. \cite{wang2016multi} researches and user-based and item-based collaborative filtering algorithms and our proposed method}\label{Fig:ADiversity}
	\end{minipage}\hfill
\end{center}
\end{figure}

In this section, the results are compared and analyzed based on the criteria introduced in previous section. Figure \ref{Fig:precision}, \ref{Fig:novelty}, and \ref{Fig:ADiversity} show a comparison of these results.
According to Figure \ref{Fig:precision}, the precision obtained by the proposed method in this study shows a higher value than other studies. Although the method proposed by Wong et al. \cite{wang2016multi} acquired 89\% for precision criterion, its precision is 2\% less than our proposed method. In this study, since we consider other users' preferences of the same age, our method can create lists with better precision. Thus, we can use more long-tail items that obtained high ratings but had limited ratings.

In Figure \ref{Fig:novelty}, again, the research of Wang et al. \cite{wang2016multi} gained a higher value for the novelty criterion. However, the proposed method in this study has obtained higher values for the novelty criterion. In the proposed method, since one of the objectives of the multi-objective problem is to use long-tail items based on the user's activity, the proposed method tries to recommend long-tail items more than short head items. With this approach, we create recommendation lists in a more personalized way. As users' interests change over time, this method can personalize recommendation lists more accuratly. Therefore, users who have been in the system for a longer time have a better opportunity to see long-tail items in their recommendation lists, which improves long-term novelty criteria.

In Figure \ref{Fig:ADiversity} again, the proposed algorithm performs better in the aggregate diversity criterion. Among other algorithms selected to compare the results, Malekzadeh et al. \cite{hamedani2019recommending} performed better, As increasing the diversity of recommendation lists is considered as one of the goals of the multi-objective algorithm. However, in terms of this criterion, it has not been as successful as the proposed method of this study. By using Equation \ref{eqPosibility}, we try to use items that were not in the previous recommendation lists. This allows us to use a wide range of items. Therefore, in the long term, items that were not in the recommendation lists will be recommended to users, and the aggregate diversity criterion will subsequently improve.

\section{Conclusion}
This study mitigates the long tail problem according to the users' dynamic preferences that change over time. The multi-objective memetic evolution algorithm has been used for this problem. In the first objective, we try not to use more popular items in the recommendation lists. In the second objective, we prevent the reduction of accuracy of the recommendation lists as much as possible. The third objective uses short-head and long-tail items based on user activity over time and user age. In the last one, since long-tail items usually received lower ratings from users, finding similarities based on ratings is not a good solution. Thus, in the last goal, recommendation lists are created based on the relationship between user age and genres preferred by users of the same age.
Besides, in the Memetic algorithm, we add items that have received high average ratings and a low number of ratings to initial populations. By doing so, long-tail items that did not appear much in the recommendation lists with the baseline algorithms will have the opportunity to appear in the users' recommendation lists.

In the future, the baseline algorithms can be changed so that the ratings recorded by users in the past are less weighted in current predicting the ratings of items. For example, if the user abandons the recommender system for a long time, his / her past interests can be changed. Items that become a trend and receive high ratings can also be used in recommendation lists to form the initial population in evolutionary algorithms. Also, items that had no ratings in the past but are very similar to items that are popular today in various respects (such as genre, synopsis) can be included in user recommendation lists.

	\section*{Declarations of interest}
	None.
	\bibliography{References} 
	
\end{document}